\documentclass[14pt,letterpaper,times]{article}
\usepackage[tcidvi]{graphicx}
\usepackage[english]{babel}
\usepackage{amsmath,amsfonts,graphics,color,colortbl,fontsmpl,amssymb,amsfonts,epsfig,rotating,float,amscd,multicol,layout}

\begin{document}
\title{Generalized concurrence and limits of separability for two qutrits}
\author{C\'esar Herre\~no-Fierro\footnote \quad , and Jagdish R. Luthra\footnote.\\
\small{Departamento de F\'isica, Universidad de los Andes, A.A.
4976 Bogot\'a, Colombia}\\
\vspace{.7cm} \small{e-mail: ($*$) c-herren@uniandes.edu.co,
($\dagger$) jluthra@uniandes.edu.co}
}
\twocolumn[
\begin{@twocolumnfalse}
\maketitle
\begin{abstract}
We present an extension of the Wootters concurrence \cite{wootters} for the case
of two qutrits in mixed states. The reduction of our extension to
the case of two levels shows complete agreement with  Wootters
concurrence for two qubits. As an explicit example, we compute the
concurrence for a family of symmetric states and we obtain the
bounds on the limit for separability. Our results are compared with
those of the negativity.\\
\end{abstract}
\end{@twocolumnfalse}
]
\section{\small{Introduction}}
Quantum entanglement is of great current interest because of its
key role in the modern theory of quantum information
\cite{chuang,ekert}. Quantum entanglement of bipartite pure states
is well understood, including the general case of arbitrary
$n$-dimensional bipartite systems. Several interesting measures of
entanglement have emerged and are in agreement with the definition
of reduced entropy given by the von Neumann entropy of the reduced
density matrix \cite{bennett}. For the case of mixed states some
measures have also emerged but most of them are not easily
computable for a general density matrix. The exceptions are the
concurrence of Wootters (and closed expression for the
entanglement of formation, EOF) \cite{wootters} for two qubits and
the generalized negativity of Vidal \cite{vidal} for an arbitrary
pair of qunits. However, Vidal's negativity is defined on the
basis of the Peres-Horodocki \cite{peres} criteria which is found
to be a necessary but not sufficient condition for separability in
higher dimensional systems than $2\times 3$ \cite{horodecki},
which implies that these measures might not detect some entangled
states with positive partial transpose (PPT). We illustrate this
point with some examples in this paper (Sec. 3).

An alternative interpretation of the role of the Pauli matrix
$\sigma_y$ is made in Sec. 1 to deduce the extended version of the
concurrence transformation for two qutrits.  In Sec. 2, we
introduce the concurrence for two qutrits and analyze the results
for the case of a general pure qutrit. In this section we also
explore the reduction of this result to the case of two level
systems and then compare them with the well known results for two
qubits.  In Sec. 3, we introduce the general definition of our
concurrence for the case of mixed states, explore the concurrence
for the case of a Werner type state and obtain the bonds on the
limit for separability for this case, and then compare our results
with those of Vidal \cite{vidal}. Finally, in Sec. 4 we present
our conclusions.
\section{\small{Concurrence transformation}}
Concurrence for two qubits systems is obtained by means of a
composition of the complex conjugate and the transformation
performed by $\sigma_y$.  The action of $\sigma$ on a single qubit
$|\psi_2\rangle=\sum_{i=0,1}c_i|i\rangle$ can be understood in a
general form as a transformation which {\it kills} the original
state and generates a new state $|\psi_2'\rangle$. The new state
is orthogonal to $|\psi_2\rangle$ if and only if $|\psi_2\rangle$
is maximally nonuniform in the $c_i$, i.e., if, $|\psi_2\rangle$
is one of the elements of the basis. Otherwise, the generated
state $|\psi_2'\rangle$ is not orthogonal to $|\psi_2\rangle$,
and, in a special case, is collinear to $|\psi_2\rangle$, if and
only if, it is made of states with equal weights of the elements
in the basis. Since $\sigma_y$ has other special properties like
hermiticity, unitarity, a square root of identity, and to has
diagonal elements equal to zero, one is motivated to search for an
extended version of $\sigma_y$ for qutrits which fulfils all of
these properties. However, it can be shown that all these
properties can not be fulfilled simultaneously by one operator in
higher dimensional systems. This arises from the fact that the
symmetry of a two-level systems is, in many ways, exceptional. We
focus our search for a silmilar transformation fulfilling not all
but the fundamental properties of $\sigma_y$ in order to measure
the concurrence of a pair of qutrits. We look for a hermitian
operator as a basis of our generalized concurrence so that the
transformation might be associated to a physical quantity, to have
all diagonal elements equal to zero so that the transformation
{\it kills} the original state, and instead of a flip or inverter
operator we will say, to be a {\it split-level} operator which is
a more general class of transformation. We define {\it
split-level} operator $S_{n,j}$ as a transformations that {\it
splits} the $j$-th component of the n-dimensional ket-vector state
$|\psi\rangle$ into a superposition of $n$-1 orthogonal
components. The generalized version of $\sigma_y$ can be obtain by
considering the well known connection between $\sigma_y$ and the
ladder operators $J_{\pm}$ \cite{chinos}
    \begin{equation}\label{sigma-ladder}
    \sigma_y=-\frac i\hbar\left(J_+-J_-\right)
    \end{equation}
for two-level systems, the ladder operator can be written in the
matrix representation as
\begin{equation*}
J_+=\left(\begin{array}{cc}
      0 & 0 \\
      1 & 0 \\
    \end{array}%
    \right)\hspace{0.5cm}\text{and}\hspace{0.5cm}J_-=\left(%
    \begin{array}{cc}
      0 & 1 \\
      0 & 0 \\
    \end{array}%
    \right)
\end{equation*}
Then Eq. (\ref{sigma-ladder}) can be written as $\sigma_y=-\frac
1\hbar \mathcal O_2$, where
    \begin{equation}\label{O2-ladder-matrix}
    \mathcal O_2=e^{i\pi/2}\left(%
    \begin{array}{cc}
      0 & 0 \\
      1 & 0 \\
    \end{array}%
    \right)+e^{3i\pi/2}\left(%
    \begin{array}{cc}
      0 & 1 \\
      0 & 0 \\
    \end{array}%
    \right)
    \end{equation}
This picture of ladder operators can be thought as a reduction of
a more general {\it split-level} operators. Furthermore, as can be
seen clearly from this picture, $J_+$ and $J_-$
define the set of all possible {\it split-level} operators for
two-level systems, and $\sigma_y$ is the sum of all possible
operators for two level systems with suitable relative phases. It
is then straightforward that for three-level systems the set of
{\it split-level} operators $S_{3,i}$ are given, in the matrix
representation, by
    \begin{equation}\label{splitters-3}
    \begin{array}{cc}
      S_{3,1}=\left(%
    \begin{array}{ccc}
      0 & 0 & 0 \\
      1 & 0 & 0 \\
      1 & 0 & 0 \\
    \end{array}%
    \right)& S_{3,2}=\left(%
    \begin{array}{ccc}
      0 & 1 & 0 \\
      0 & 0 & 0 \\
      0 & 1 & 0 \\
    \end{array}%
    \right)
    \end{array}
    \end{equation}
    \begin{equation*}
    S_{3,3}=\left(%
    \begin{array}{ccc}
      0 & 0 & 1 \\
      0 & 0 & 1 \\
      0 & 0 & 0 \\
    \end{array}%
    \right)
    \end{equation*}
The corresponding operator for two qutrits can be initially
defined as
    \begin{equation}\label{O3'}
    \mathcal O_3'=e^{i\pi/3}S_{3,1}+e^{3i\pi/3}S_{3,2}+e^{5i\pi/3}S_{3,3}
    \end{equation}
Although the relative phases have been introduced following the
same sense as that of the case of qubits (equally spaced along the
whole phase space), the operator $\mathcal O_3'$ is not hermitian
as it should be. These equally spaced phases in general do not
lead to hermitian operators even in the case of even-dimensional
systems, different from two-level systems. This is again a feature
that only occurs in the exceptional dimension of two-level
systems.

Then, we will say, instead of equally spaced relative phases
between {\it split-level} operators, that the relative phases are
to be uniformly and maximally spaced in one semi-plane so that the
hermiticity of $\mathcal O$ can be granted. These relative phases
can be obtained for three level systems by introducing the set of
diagonal matrices $\{H_{3,i}\}$
    \begin{equation}\label{H}
    \begin{array}{cc}
      H_{3,1}=\left(%
    \begin{array}{ccc}
      -1 & 0 & 0 \\
      0 & i & 0 \\
      0 & 0 & -i \\
    \end{array}%
    \right) & H_{3,2}=\left(%
    \begin{array}{ccc}
      -i & 0 & 0 \\
      0 & -1 & 0 \\
      0 & 0 & i \\
    \end{array}%
    \right)
        \end{array}
    \end{equation}
    \begin{equation*}
     H_{3,3}=\left(%
    \begin{array}{ccc}
      i & 0 & 0 \\
      0 & -i & 0 \\
      0 & 0 & -1 \\
    \end{array}%
    \right)
\end{equation*}
which can be generated by permuting the diagonal entries obtained
from rotation of $\pi/2$ in the phase space starting from $\pi/2$,
i.e., $e^{i\pi/2}$, $e^{2i\pi/2}$, and $e^{3i\pi/2}$. Now if we
define the vector $\mathbf{H}=H_{3,1}\hat u_1+H_{3,2}\hat
u_2+H_{3,3}\hat u_3$, where the $\hat u_i$ are orthogonal unitary
vectors, and the vector $\mathbf{S_3}=S_{3,1}\hat u_1+S_{3,2}\hat
u_2+S_{3,3}\hat u_3$, then the transformation for qutrits can be
redefined as
    \begin{equation}\label{O3-vector}
        \mathcal O_3=\mathbf H_3\cdot\mathbf S_3
    \end{equation}
In the matrix representation it reads
    \begin{equation}\label{O3-matrix}
    \mathcal O_3=\left(%
    \begin{array}{ccc}
      0 & -i & i \\
      i & 0 & -i \\
      -i & i & 0 \\
    \end{array}%
    \right)
    \end{equation}
In terms of an alternate convenient decomposition, $\mathcal O_3$
can be written as
    \begin{equation}
    \label{O3-X-Xdagger}
    \mathcal{O}_3=i\left(X-X^{\dagger}\right)
    \end{equation}
where $X$ is the cyclic (but not hermitian) Pauli matrix for
qutrits. A complete characterization of the Pauli matrices for
qutrits can be found in Ref. \cite{Lawrence}.

This decomposition suggest a simpler interpretation of $\mathcal
O_3$.  Since $X$ and $X^{\dagger}$ are the cyclic ladder operators
for two qutrits ($X$ increases and $X^{\dagger}$ decreases,
cyclicly the level of a single qutrit), we can extend the
definition of $\mathcal O_3$ to a general case of qunits as the
composition of all possible cyclic operator with suitable relative
phases.

Summarizing the characteristics of $\mathcal O_3$, we have that it
is hermitian, non-unitary, and its action on a single qutrit can
be described as follows
    \begin{equation}\label{operator-action}
    \mathcal O|j\rangle=i(|j+1\rangle-|j-1\rangle)
    \end{equation}
where $j\pm1$ is modulo 3.
\section{\small{Concurrence for two qutrits in a pure state }}
Now we extend the concurrence of Wootters by defining the
concurrence for two qutrits in pure state
$|\psi_{3\times3}\rangle=|\psi\rangle$ as
\begin{equation}\label{conc-3-pure1}
    C_3(\psi)=|\langle\psi|\tilde\psi\rangle|
\end{equation}
where
    \begin{equation}\label{psi-tilde}
        |\tilde\psi\rangle=(\mathcal O_3\otimes\mathcal
        O_3)|\psi^*\rangle
    \end{equation}
with $\mathcal O_3$ as defined in (\ref{O3-X-Xdagger}), and
$|\psi^*\rangle$ being the complex conjugate of $|\psi\rangle$.
For a vector state of two qutrits, in the standard basis
$\{|i,j\rangle\}$,
    \begin{equation}\label{general-qutrit}
        |\psi\rangle = \sum_{i,j=0}^2\alpha_{ij}|i,j\rangle
    \end{equation}
with, $ \sum_{i,j=0}^2\alpha_{ij}^2=1$, concurrence
(\ref{conc-3-pure1}) can be written as
  \begin{align}\label{conc-gen-state}
        C_3(\psi)&=\\
        &|(\alpha_{00}+\alpha_{11}+\alpha_{22})^2+(\alpha_{01}+\alpha_{12}+\alpha_{20})^2\nonumber\\
        +&(\alpha_{02}+\alpha_{21}+\alpha_{10})^2-(\alpha_{00}+\alpha_{12}+\alpha_{21})^2\nonumber\\
        -&(\alpha_{01}+\alpha_{10}+\alpha_{22})^2-(\alpha_{02}+\alpha_{11}+\alpha_{20})^2|\nonumber
    \end{align}
In the Schmidt basis $\{|i,i\rangle\}$, state
(\ref{general-qutrit}) can be decomposed as
    \begin{equation}\label{psi-schmidt}
    |\psi\rangle = \sum_{i=0}^2\beta_i|i,i\rangle
    \end{equation}
and then (\ref{conc-3-pure1}) reduces to
    \begin{equation}\label{conc-schmidt-qutrit}
        C_3(\psi)=\left|\left(\sum_{i=0}^2\beta_i\right)^2-1\right|
    \end{equation}
    where the $\beta_i$ are the Schmidt coefficients.

It is remarkable that results (\ref{conc-gen-state}) and
(\ref{conc-schmidt-qutrit}) include the lower dimensional case of
two qubits. The corresponding results for two qubits can be
recovered by setting to zero all the coefficients within the level
to be eliminated. For instance, if we set to zero all the
$\alpha_{i2}$ and $\alpha_{2j}$ in (\ref{conc-gen-state}), and
consequently, the Schmidt coefficient $\beta_{2}$ in
(\ref{conc-schmidt-qutrit}), then, the new expressions coincide
with those for two qubits, namely,
    \begin{equation}\label{con-woo-coef}
            C_2(\psi)=2|\alpha_{00}\alpha_{11}-\alpha_{01}\alpha_{10}|
    \end{equation}
and, respectively,
    \begin{equation}\label{con-schmidt-coef}
            C_2(\psi)=2\beta_{0}\beta_1
    \end{equation}
$C_3$ predicts an amount of entanglement of 2 for maximally
entangled states against the value 1 for the case of two qubits.
This is an expected result, since the amount of entanglement
should increase with the dimensionality of the system. This might
be associated with the possibility of two different flip action in
three level systems against only one flip in two levels, and
suggests than in the general case of two qunits (n-level
particles) the concurrence of maximally entangled states is given
by
        \begin{equation}\label{Cn-max}
                C_{n,\text{max}}=n-1
        \end{equation}

By comparing our results with those of G. Vidal we find exact
agreement between one of his negativities, the so called {\it
robustness} ($\mathcal N_{\mathcal{SS}}$). The negativity is
defined as the sum of the negative eigenvalues of the partial
transpose of the density matrix, so that it measures the degree to
which the partial trace of the density matrix ($\rho^{T_A}$) fails to
be positive.  It is given by (Eq. (1) in Ref. \cite{vidal})
        \begin{equation}\label{negativity}
        \mathcal{N}=\frac{\|\rho^{T_A}\|-1}2
        \end{equation}
Where $\|.\|$ is the trace norm. For the case of pure states, the
negativity $\mathcal N$ (Eq. (47) in Ref. \cite{vidal}) reduces
exactly to half of Eq. (\ref{conc-schmidt-qutrit}) and then,
robustness ($\mathcal N_{\mathcal{SS}}=2\mathcal N$) matches
exactly our result. This result gives a prescription to generate
negativity by a procedure similar to that of the concurrence given
by Pauli spin operator.
    \begin{figure}
    \begin{center}
    \includegraphics[angle=270,width=6.4cm]{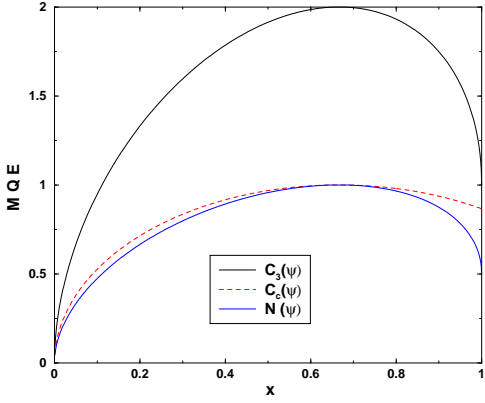}
    \end{center}
    \caption{Measures of Quantum Entanglement (MQE) for two
    qutrits in pure state (\ref{psi-schmidt}). Concurrence $C_3(\psi)$
    (\ref{conc-schmidt-qutrit}), and therefore robustness $\mathcal{N_{SS}}$, in the solid upper line, Cereceda's
    concurrence $C_c(\psi)$ (\ref{cereceda-conc-smith}) in the dashed
    line, and negativity $\mathcal N$ (\ref{negativity}) in the solid
    lower line, against the parameter $x$, with
    $\beta_0=\beta_1=\sqrt{x/2}$ and $\beta_2=\sqrt{(1-x)}$}\label{PlotMQEpure}
    \end{figure}

On the other hand, J. Cereceda \cite{cereceda} has obtained an
extension of concurrence for two qutrits systems by means of
decomposing the standard density matrix into the SU(3) generators.
His concurrence ($C_c$) for the case of two qutrits in pure state
(\ref{psi-schmidt}) is given by (Eq. (21) in Ref. \cite{cereceda})
    \begin{equation}\label{cereceda-conc-smith}
    C_c(\psi)=\sqrt{3\left((\beta_0\beta_1)^2+(\beta_1\beta_2)^2+(\beta_2\beta_0)^2\right)}
    \end{equation}
This result does not match ours.  Besides, $C_c$ is not a general
result extendable for mixed states, and apart from its
normalization drawback, $C_c$ does not yield the expected values
for the case of two qubits. For example, it predicts a value of
concurrence $\frac{\sqrt3}2$ for partially entangled qutrits
states against 1, as it should be since the amount of entanglement
of these two partially entangled qutrits is the same as that of
two maximally entangled qubits.
\section{\small{Concurrence of two qutrits in a mixed state}}
For the case of a mixed state
$\rho=\sum_ip_i|\psi_i\rangle\langle\psi_i|$ of two qutrits,
concurrence is defined as the average concurrence of the pure
states making up $\rho$ minimized over all possible decompositions
of $\rho$. That is
\begin{equation}\label{conc-mixed}
    C_3(\rho)=\text{min}_d\overline{C_3(|\psi\rangle)}=\text{min}_d\sum_ip_iC_3(|\psi_i\rangle)
\end{equation}
where, min$_d$ means the minimization over all possible
decompositions of $\rho$ and $C_3(|\psi_i\rangle)$ is the
concurrence of the $i$-th pure state in the ensemble that defines
the state $\rho$.

For two qutrits in a mixed state the concurrence is then given by
\begin{equation}\label{conc-mixed-lambda}
C_3(\rho)=\text{max}\{0,2\lambda_1-\sum_{i=1}^9\lambda_i\}
\end{equation}
where the $\lambda_i$ (with $i=1,2,...,9$) are the square roots of
the eigenvalues of the non-Hermitian matrix $\rho\tilde{\rho}$ in
decreasing order, where
\begin{equation}\label{rhotilde}
\tilde\rho=(\mathcal O_3\otimes\mathcal O_3)\rho^*(\mathcal
O_3\otimes\mathcal O_3)
\end{equation}
with $\rho^*$ being the complex conjugate of $\rho$. The result
(\ref{conc-mixed-lambda}) is a conjecture at this stage. The
motivation for this is the equivalence obtained and proved for the
case of two qubit systems by Wootters \cite{wootters}. In the
following section we explore this result for the case of a
standard Werner state, which allowed us to gain some confidence in
order to presume the validity of this result.
\subsection{\small{An Explicit Example: Standard Werner states}}
In two level systems a Werner state is defined as a composition of
one fraction ($x$) of a singlet ($\Psi_{\pm}$, $\Phi_{\pm}$) and
one fraction $(1-x)$ of the maximally mixed density matrix
($\frac14\delta_{i\mu,j\nu}$). For two qutrits, we consider
standard Werner kind of states as a single fraction of maximally
entangled qutrits and a random fraction of impurity (decoherence),
as follows
\begin{equation}\label{werner-state}
\rho_w=\frac x3\sum_{i,j=0}^2|i,i\rangle\langle
j,j|+\frac{1-x}9\mathbb{I}
\end{equation}
where $\mathbb{I}$ is the identity matrix in the Schmidt basis.
For this case, the nine eigenvalues of $\rho\tilde{\rho}$ are

\begin{equation}
\begin{array}{ccc}\label{eigenvaluesRotilde}
0 & &\text{of multiplicity 5}\\
\frac1{9}\left(1 - x\right)^2 & &\text{of multiplicity 3}\\
\frac1{9}\left(1 + 5x\right)^2& &
\end{array}
\end{equation}
among which the last one is the greatest for $0\leq x\leq 1$, and
then, concurrence $C_3$ (\ref{conc-mixed-lambda}) is given by
\begin{equation}
C_3(\rho_w)=\text{max}\{ 0,\frac 83x-\frac 23\}
\end{equation}
which yields that our measure is zero for $x\leq \frac14$. For
$x>\frac14$ concurrence (\ref{conc-mixed-lambda}) increases
linearly with $x$, up to its maximum value 2, for $x=1$, as
expected for maximally entangled qutrits.

From this result we can recover the results for two-level systems
by setting some of the entries in the density matrix to be zero.
Let us drop the third level $|2\rangle$ by setting the Schmidt
coefficients $\beta_0=\beta_1=\frac12$, and $\beta_2=0$ in the
definition of the maximally entangled state (now, actually partial
entangled qutrit) for the standard Werner state
(\ref{werner-state}), so that
$|\psi_{max}'\rangle=\sum_{i=0}^2(1-\delta_{i,2})(1-\delta_{2,j})\beta_i|i,i\rangle=\sum_{i=0}^1\beta_i|i,i\rangle$.
Also we have to drop the elements $\mathbb{I}_{2\mu,j\nu}$,
$\mathbb{I}_{i2,j\nu}$, $\mathbb{I}_{i\mu,2\nu}$ and
$\mathbb{I}_{i\mu,j2}$ in the maximally mixed density matrix, so
that the elements in maximally mixed component of the standard
Werner state would read
$\frac19(1-\delta_{2\mu,j\nu})(1-\delta_{i2,j\nu})(1-\delta_{i\mu,2\nu})
(1-\delta_{i\mu,j2})\delta_{i\mu,j\nu}=\frac14\delta_{i\mu,j\nu}(1-\delta_{i\mu
j\nu,2})$. Meanwhile computing the concurrence for this particular
case we obtain four eigenvalues of $\rho'\tilde{\rho'}$ different
to zero, namely, $\frac1{16}\left(1 - x\right)^2$ (of multiplicity
2), and $\frac1{16}\left(1 + 12x - 5x^2 \pm 4x\sqrt{1 + 12x -
9x^2}\right)$. Then, the concurrence is given by
    \begin{equation}
        C_{3\rightarrow2}(\rho_w)=\text{max}\{0,c\}
    \end{equation}
with
\begin{align*}
c=&\frac14\sqrt{1 + 12x - 5x^2 + 4x\sqrt{1 + 12x -
    9x^2}}\\
    -&\frac14\sqrt{1 + 12x - 5x^2 - 4x\sqrt{1 + 12x -9x^2}}-\frac12\left(1 - x\right)
\end{align*}
This implies that this density matrix becomes separable for all
$c\leq 0$, i.e., $x\leq\frac13$ which is exactly the prescribed
value by Peres \cite{peres} for two qubit systems.  On the other
hand, the maximum concurrence for this case is 1, which also
coincides with the results for maximally entangled qubits, as
expected.

For standard Werner states (\ref{werner-state}) Vidal's negativity
and robustness (expressions (51) and (52) in Ref. \cite{vidal},
with $d=3$, and $g=0$ which are the dimension of the two qudits,
and a parameter used to consider a more general symmetric state,
respectively) read
\begin{equation}\label{neg}
\mathcal N(\rho_w)=\frac14\left(|1-3x|+|1+3x|\right)-\frac12
\end{equation}
and
\begin{equation}\label{robustness}
\mathcal
N_{\mathcal{SS}}(\rho_w)=\frac12\left||6x-1|-1\right|=2\mathcal
N(\rho)
\end{equation}
respectively. These two measure have maximum values of 1 and 2,
respectively, but according to them, for $x\leq\frac13$, $\rho$
becomes separable while our measure indicates that it occurs for
values of $x\leq \frac14$. This implies that our measure does
detect some entangled states that neither negativity nor
robustness do.  In fact, negativities do not make difference
between separability in two qubits and two qutrits. This can be
explained because negativities are defined on the basis of the
Peres-Horodocki criteria and, as mentioned in the introduction,
the PPT is a necessary but not sufficient condition for higher
dimension than $2\times3$ which implies that the negativities do
not detect some PPT entangled states in these
cases which include the case of the present work.
\begin{figure}
\begin{center}
\includegraphics[angle=270,width=6.4cm]{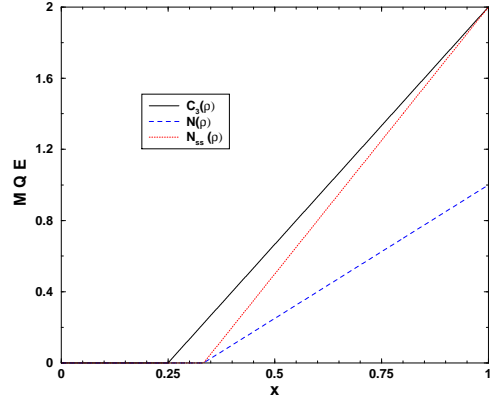}
\end{center}
\label{MQE-mixed} \caption{MQE
        of two qutrits in standard Werner state (\ref{werner-state}),
        against the parameter $x$. The solid line is the concurrence
        $C_3(\rho)$ given by (\ref{conc-mixed-lambda}), the dashed
        line is the negativity $\mathcal N(\rho)$ given by
        (\ref{neg}) and the dotted line is the robustness $\mathcal{N_{SS}}$ given by (\ref{robustness}).}
\end{figure}

Other explicit examples, as well as some guide lines to the
extension of this work to the case of arbitrary dimensional
systems will be presented in further papers \cite{tobepublished,
tobepublished2}.
\section{\small{Conclusions}}
We have generalized the Wootters's definition of concurrence for
two-level bipartite systems.  We identify a composite
transformation to the original state which includes a complex
conjugation of the state and a flip action performed by the Pauli
matrix $\sigma_y$. Starting from this analysis we generalize the
characteristics for a good ``{\it flip}'' operator in three-level
bipartite systems which leads us to a unique operator $\mathcal
O_3$ which does not have all the properties of $\sigma_y$ but the
three fundamental one: to be a composition of {\it split-level}
operators, to have null diagonal elements, and hermiticity.  Then
we define concurrence $C_3$ for two qutrits systems on the basis
of the transformation given by the operator $\mathcal O_3$ and the
complex conjugate. This definition lead us to a very well behaved
measure of entanglement for pure states.  For mixed states we
presume that the average concurrence of the pure states in the
decomposition of the density matrix $\rho$, minimized over all
possible decompositions, reduces in the same way that Wootters
concurrence reduces for two qubit systems in terms of the
eigenvalues of $\rho\tilde\rho$. We explored this measure on
explicit examples and some of the remarkable results are: 1.
Concurrence in two qutrits is stronger than in two qubits, 2. The
extended concurrence can be reduced to the case of two qubits
recovering exactly the same well known results for this case, 3.
Separability for a density matrix with single fraction of
maximally entangled qutrits and a random fraction of impurity
(decoherence) occurs for a higher random fraction of impurity that
for the case of Werner states of two qubits, namely, our results
shows that the needed random fraction of impurity in the density
matrix to be separable is $\frac34$, against $\frac 23$ in the
case of two qubits, which is consistent with the previous results,
4. Compared with other measures, ours shows some differences that
enhance its convenience, namely, regarding the close measures of
G. Vidal, the results show that those measures do not detect some
entangled states of two qutrits that our concurrence does. This
result finds foundations in the fact that those measures have been
built on basis of the Peres cirteria which is not a sufficient
condition for separability in higher systems than $2\times3$.

These results suggest that our presumption about the analytic
reduction of concurrence for mixed states is valid and that our
operator $\mathcal O_3$ gives a prescription to generate
negativity by a procedure similar to that of the concurrence given
by Pauli spin operator. It is also remarkable that our concurrence
is an effectively computable measure and seems to be extendable to
the case of $n\times n$ quantum systems as it will be presented in
a further paper \cite{tobepublished2}.

The main contribution of this work is to generate a effectively
computable measure of quantum entanglement for the case of qutrit
bipartite systems which consists on the first step toward the
definition of a general measure of quantum entanglement for two
qunits ($n$-level particles) systems. This work is also, in
general, a contribution toward the characterization and
conceptualization of the quantum entanglement.\newpage


\begin{thebibliography}{llllll}

\bibitem{wootters}W.K. Wootters, Phys. Rev.
Lett. \textbf{80}, 2245 (1998); S. Hill and W.K. Wootters, ibid.
\textbf{78}, 5022 (1997).

\bibitem{chuang}M. Nielsen, and I.
Chuang \textit{Quantum Computation and Quantum Information},
Cambridge University Press, (New York, USA, 2000).

\bibitem{ekert}A. Ekert, and A. Zeilinger \textit{The Physics of Quantum
Information}, (Springer, Berlin, 2000).

\bibitem{bennett} C.H. Bennett, H.J. Bernstein, S. Popescu, and B. Shumacher, Phys. Rev. {\bf 53}, 2046 (1996).

\bibitem{vidal}G. Vidal and R.F. Werner, Phys. Rev. A.
\textbf{65}, 032314 (2002).

\bibitem{peres} A. Peres, Phys. Rev. Lett. \textbf{77}, 1413,
(1996).

\bibitem{horodecki} M. Horodecki, P. Horodecki, and R.
Horodecki, Phys. Lett. A \textbf{223}, 1 (1996).

\bibitem{chinos}Y. Li and G. Zhu,
preprint, quant-ph/0308139 v1, (2003).

\bibitem{Lawrence} J. Lawrence, Phys. Rev. A.\textbf{70}, 012312 (2004).

\bibitem{cereceda} J.L. Cereceda, quant-ph/0305043, (2003).

\bibitem{tobepublished}C. Herre\~no-Fierro, and J. R. Luthra: ``Lower bound for separability in symetric qutrit
states''. To be published.

\bibitem{tobepublished2}C. Herre\~no-Fierro, and J. R. Luthra: ``Concurrence for arbitrary dimesional bipartite systems''. To be published.

\end{thebibliography}
\end{document}